\documentclass[journal]{IEEEtran}

\usepackage{textcomp}
\usepackage{ccaption}
\usepackage{amsmath, amssymb}
\usepackage{ upgreek }
\usepackage{bm}
\usepackage{algorithm}
\usepackage{algorithmicx}
\usepackage{algpseudocode}
\usepackage{cite}
\usepackage{graphicx}
\usepackage{graphics}
\usepackage{adjustbox}
\usepackage{subfig}
 \usepackage{booktabs}

\usepackage{subfig}
\usepackage{multirow}
\usepackage{float}
\usepackage[para]{threeparttable}
\usepackage{booktabs}
\usepackage{hyperref}
 \usepackage[outercaption]{sidecap}    

\pdfoutput=1
\begin{document}
\title{Real-Time Predictive Control Strategy Optimization}

\author{Samarth~Gupta,~
        Ravi~Seshadri,
        Bilge~Atasoy,
        A. Arun~Prakash,
        Francisco~Pereira,
        Gary~Tan,
        and~Moshe~Ben-Akiva
        

\thanks{S. Gupta, A. Prakash  and M. Ben-Akiva are with Massachusetts Institute of Technology (MIT). \{samarthg, arunprak , mba\}@mit.edu}
\thanks{R. Seshadri is with Singapore-MIT Alliance for Research and Technology (SMART). ravi@smart.mit.edu }
\thanks{B. Atasoy  is with Delft University of Technology (TU Delft). b.atasoy@tudelft.nl}
\thanks{G. Tan is with the National University of Singapore (NUS). gtan@comp.nus.edu.sg}
\thanks{F. Pereira is with the Technical University of Denmark. camara@dtu.dk}

}

\maketitle

\begin{abstract}
%
Traffic congestion has lead to an increasing emphasis on management measures for a more efficient utilization of existing infrastructure. In this context, this paper proposes a novel framework that integrates real-time optimization of control strategies (tolls, ramp metering rates, etc.) with guidance generation using predicted network states for Dynamic Traffic Assignment systems. The efficacy of the framework is demonstrated through a fixed demand dynamic toll optimization problem which is formulated as a non-linear program to minimize predicted network travel times. A scalable efficient genetic algorithm is applied to solve this problem that exploits parallel computing. Experiments using a closed-loop approach are conducted on a large scale road network in Singapore to investigate the performance of the proposed methodology. The results indicate significant improvements in network wide travel time of up to 9\% with real-time computational performance.
\end{abstract}

\begin{IEEEkeywords}
dynamic toll optimization, dynamic traffic assignment (DTA), predictive control optimization, large-scale network, real-time traffic management
\end{IEEEkeywords}

\section{Introduction}\label{sec:Introduction}
.

\IEEEPARstart{U}{rban} transportation networks are subject to large degree of variability due to the fluctuating supply and demand characteristics. These fluctuations result in the pervasive phenomena of recurrent and non-recurrent congestion, which is an escalating problem worldwide. The adverse impacts of the resulting congestion include high travel delays, high travel costs, and significant costs to the economy and environment. Consequently, there has been an increased emphasis on developing tools to mitigate congestion and efficiently utilize existing infrastructure. In this context, we propose an integrated framework ---within a Dynamic Traffic Assignment (DTA) system--- to optimize network control strategies in real-time considering network state predictions. Specifically, the generated control strategies are predictive (or proactive) as opposed to being just reactive. The framework also incorporates the generation of consistent guidance ---it ensures that the guidance
disseminated considers the travelers response to it, thereby increasing the reliability of the provided information. Further, we demonstrate the effectiveness of the proposed framework through a real-world application to the predictive optimization of network tolls. 

The motivation for this study is fourfold. First, the need for decision support tools to facilitate a more efficient utilization of existing infrastructure. Second, most studies on optimal network control do not combine the optimization of network control strategies with the generation of guidance information. The third motivating factor is the complexity and scale of the problem. As the objective function involves simulation, it tends to be non-linear and non-convex making it challenging for a real-time application. Finally, the study is also motivated by important applications in real-time traffic management and incident response systems.

In view of the aforementioned motivations, the following objectives are identified: 1) To develop an integrated framework within a real-time DTA system that determines optimal control strategies and consistent guidance information considering traffic state predictions; 2) To propose a real-time solution methodology to efficiently solve for the optimal strategies under the framework proposed in objective 1; 3) To evaluate the proposed framework using a closed-loop approach (where the DTA system is interfaced with a traffic microsimulator that emulates the stochasticity in real world, thus providing a platform for realistic evaluation) on a large real-world network with link tolls as control strategies. 

The salient contributions of this work are, first, the proposed simulation-optimization framework simultaneously optimizes network control strategies and computes consistent guidance information based on traffic state predictions. Utilizing traffic state predictions aids in accurately evaluating the effect of control strategies. Furthermore, the control strategy at any location is determined based on global traffic state predictions and not just local predictions, thereby explicitly considering the system-level effects. The consistency in guidance ensures that the information disseminated by the traffic management center is reliable, an important issue that has been overlooked in the literature on control strategy optimization. The second contribution is that we apply a highly parallelizable genetic algorithm to solve for the optimal control strategy (within the proposed framework) that maintains computational tractability to achieve real-time performance on a large real-world network. Third, we evaluate the proposed framework using a rigorous closed-loop approach that ensures that impacts of the control strategy are not overestimated. The experiments demonstrate the effectiveness of the proposed system which can yield travel time improvements of up to 9\%, and average computational times of less than 5 minutes. In addition, a sensitivity analysis is performed with respect to network demand levels.

\section{Literature Review}\label{sec:LitReview}

Although the framework presented in this paper is applicable to other control strategies including ramp-metering, the review here focuses on real-time congestion pricing in view of the application presented. The reader is referred to Chung and Recker (2011) for a review of existing toll facilities in the US \cite{Chung2011} and to de Palma and Lindsey (2011) \cite{dePalma2011} for a discussion of congestion pricing technologies.

There are two broad categories of tolling strategies: fixed pricing strategies and dynamic pricing strategies. In fixed pricing strategies, the tolls are predetermined; they can be a time-invariant or can vary in a predetermined manner during the day (time-of-day tolling). Further, in a fixed pricing strategy, tolls can also vary based on location and vehicle type. In the dynamic pricing strategies, the tolls are continually determined based on the current/future traffic conditions and are not predetermined. A dynamic tolling strategy can be either reactive or predictive. In a reactive tolling strategy, the tolls are determined based on the current traffic conditions. In contrast, in predictive tolling, the tolls are determined considering predicted traffic states. 

Yang (2005) \cite{Yang2005} and Tsekeris and Voss (2009) \cite{Tsekeris2009} should be referred for a review of work on static and fixed congestion pricing. Among the studies that determine time-dependent and fixed pricing, De Palma et al. (2005) was one of the earliest to study the effect of time-invariant vs. time-dependent pricing using a simulator \cite{dePalma2005}. Their experiments show that time-dependent tolls can generate twice the welfare gains compared to time-invariant tolls. Xu (2009) presented an optimization framework with the travel time objective and solved the problem using the SPSA (Simultaneous Perturbation Stochastic Approximation) algorithm \cite{Xu2009}. Chen et al. (2014) solve the similar problem with the travel time objective \cite{Chen2014}. The problem was solved by statistically modeling the objective function (calculated from the output of DynusT) using Kriging. The same authors later extended the work to objectives of throughput and revenue\cite{Chen2016}. The tolling scheme was based on the vehicle miles traveled. \cite{jenny} has also studied distance based tolling along with elastic demand.

The studies on the dynamic reactive pricing have predominantly been in the context of managed-lane operations. Yin and Lou (2009) propose two dynamic pricing approaches for managed toll lanes: a feedback-control approach and reactive self-learning approach \cite{Yin2009}. The pricing decisions are based on real-time traffic conditions and the objective is to improve the free-flow travel service on the toll lanes while maximizing total throughput. Similar approaches ---based on feedback control--- have been used to optimize for various other objectives like speed, travel time, delays, and revenues \cite{Zhang2008,Lou2011,Kitae2014}.
Morgul (2010) studied dynamic reactive pricing for different tolled links in a network by employing the traffic simulation software Paramics and TransModeler \cite{Morgul2010}. The algorithm applied was from Zhang et al. (2008) \cite{Zhang2008}; it is a feedback controller based on speed measurements. It was shown that dynamic tolling results in lower queue lengths and higher speeds.

Dong et al. (2011) studied the predictive tolling strategy, where the predicted traffic conditions provided by DYNASMART-X were used to generate the tolls \cite{Dong2011}. A feedback control approach was adopted where the toll at a location is determined by adjusting the previous toll based on the deviation of predicted concentration on the corresponding link from the desired level. Hassan et al. (2013)\cite{Hassan2013} also studied predictive tolling in order to maximize revenue. The toll is optimized based on a formulation where a Greenshields model is embedded to represent traffic dynamics and a binary logit model is incorporated for route choice. A linear approximation is used for the solution of the optimization model and the optimized toll is evaluated through a simulation-based DTA system (DIRECT) with prediction capabilities. They applied the tolling methodology on a synthetic corridor network with two gantries where the tolls need to be optimized. More recently,  Hashemi and Abdelghany (2016)\cite{Hashemi2016} provided a predictive control framework with an example of timing decisions on signalized intersections. They also used DIRECT for state estimation and prediction and for the optimization of the control they used genetic algorithm similar to our approach. They applied the methodology to the US-75 corridor in Dallas. 

In summary, a considerable number of studies have a reactive setting, i.e., they do not consider the effects in future time-periods while determining tolls in the current time-period. This myopic tolling policy can result in undesirable and fluctuating tolls and traffic conditions. Additionally, a common approach to determine the dynamic tolls is based on feedback control, where the tolls are adjusted based on either observed or predicted characteristics like speed or queues.  However, as the characteristics of only the tolled links are used to determine the corresponding tolls, the system-level interactions are ignored and hence makes them inefficient for large scale networks. Recent studies moved to predictive tolling as some major examples are cited above. However, few studies optimize for predictive dynamic pricing strategies at a network-level, most consider corridor type networks. Furthermore, consistency between the provided guidance and the resulting network conditions is not handled fully in most of the studies. Finally, the evaluation of the optimized tolls is done through the same simulator that is used to optimize the tolls \cite{sgupta2016}.  This may overestimate the network performance improvements. This study addresses these gaps in real-time predictive control systems, more specifically tolling.

\section{Integrated Framework for Real Time Control Strategy Optimization and  Guidance Generation}\label{sec:Framework}
\begin{figure*}[t]
\centering
{\includegraphics[scale =0.45]{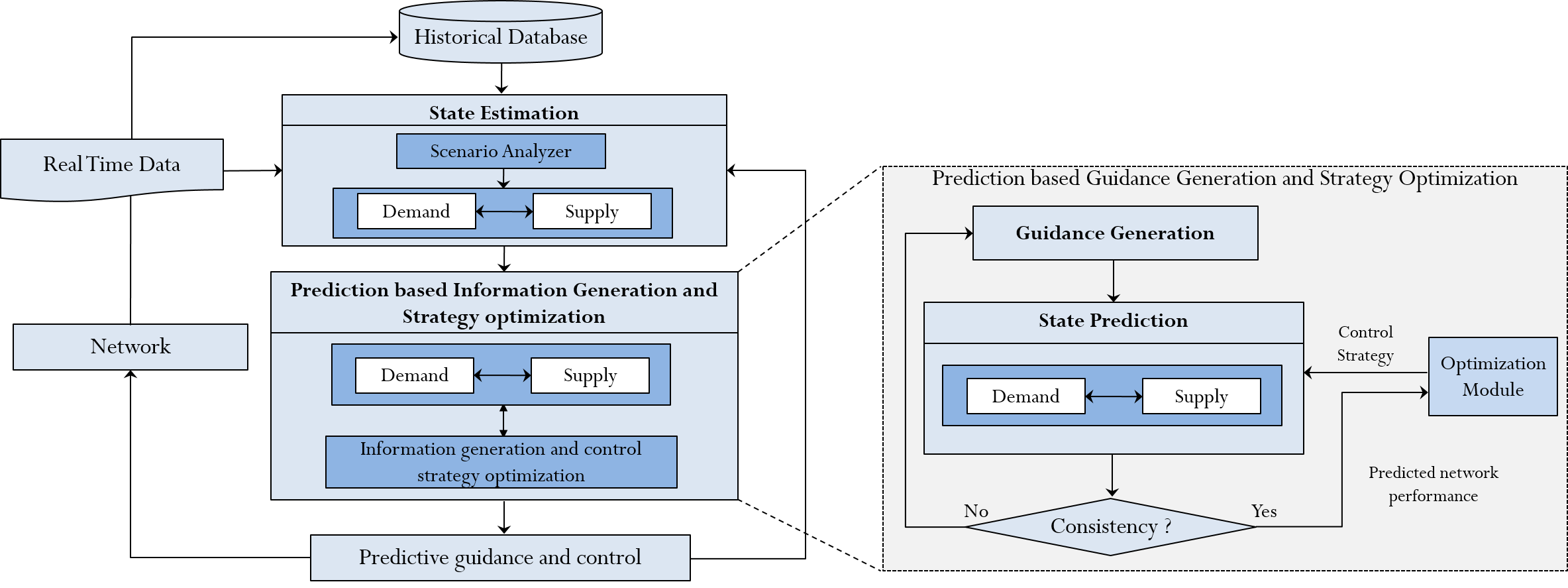}}
\caption{Framework for Integrated Guidance Generation and Control Strategy Optimization}
\label{fig:Framework}
\end{figure*}

This section briefly describes the proposed framework for the integrated optimization of control strategies and generation of consistent travel time guidance. For the ease of exposition, the framework is illustrated using DynaMIT2.0, a simulation based DTA system for traffic  state estimation and prediction developed at the MIT Intelligent Systems Laboratory \cite{BenAkiva_et_al:DynaMIT2010,7313455}. However, it is noted that the framework is generic and applies to any real-time DTA system. The DynaMIT2.0 system is first very briefly introduced followed by a discussion of the proposed framework. 

DynaMIT2.0 is composed of two core modules, state estimation and state prediction, and operates in a rolling horizon mode. During each execution cycle, the state estimation module uses a combination of historical information and real-time data from various sources (surveillance sensors, traffic information feeds, weather forecasts) to first calibrate the demand and supply parameters of the simulator so as to replicate prevailing traffic conditions as closely as possible. The updated parameters are then utilized to estimate the state of the entire network for the current  time interval. Based on this estimate of the current network state, the state prediction module predicts future traffic conditions for a prediction horizon and generates consistent guidance information (refer to \cite{BenAkiva_et_al:DynaMIT2010} for more details on the DynaMIT) that is disseminated to the travelers. 

The integrated framework is summarized in Figure \ref{fig:Framework}. During each execution cycle, following \textit{state estimation}, the \textit{Prediction based Information Generation and Strategy optimization} process is invoked. Within this process, the optimization module generates a series of control strategies (for example network tolls, signal timings, etc.) for the prediction horizon period which are to be evaluated on the basis of a specific objective. This can include the minimization of total system travel time, maximization of consumer surplus, maximization of operator revenues and so on. The evaluation of each control strategy involves running the state prediction module iteratively to ensure that the predicted network state is consistent with the provided guidance. 

More specifically, the state prediction module (expanded in the right half of Figure \ref{fig:Framework}) begins with the most recently disseminated guidance (for instance, the guidance may be in the form of network link travel times) as a trial solution. 
The coupled demand and supply simulators are then used to predict the network state based on the given control strategy and assumed guidance as inputs (note that the route choices of drivers change in response to the control strategy and guidance). This yields predicted network travel times which are then combined with the original guidance (using the method of successive averages or MSA) to obtain a revised travel time guidance solution. This procedure is iteratively performed until convergence, i.e., the provided travel time guidance and predicted network travel times are within a pre-specified tolerance limit $\epsilon_P$. Once convergence is achieved, the state prediction and guidance strategy are termed 'consistent' and the corresponding network state is then used by the optimization module to evaluate the objective function and search for the optimal control strategy. Following the completion of the optimization procedure, the \textit{Prediction based Information Generation and Strategy optimization} process returns an optimal control strategy that is applied to the network and consistent travel time guidance that is disseminated to travelers.

The proposed framework is demonstrated in the subsequent sections through an application to the dynamic toll optimization problem.

\section{Formulation of Dynamic Toll Optimization Problem}\label{sec:Formulation}
\begin{figure}[h]
\centering
{\includegraphics[scale =0.45]{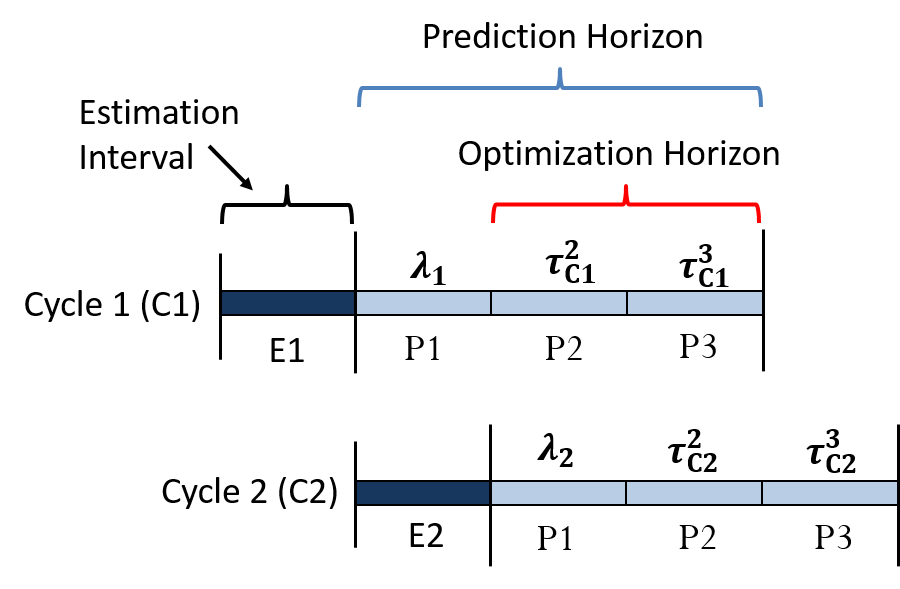}}
\caption{Illustration of the rolling horizon approach for toll optimization}
\label{fig:Formulation}
\end{figure}

The transportation network of interest is represented as a directed graph $G(N,A)$ where $N$ represents the set of $n$ network nodes and $A$ represents the set of  $m$ directed links. Let $ \widetilde{A} \subseteq A$ represent a subset of network links that are tolled with $ \widetilde{m} = | \widetilde{A} | $. Consider an arbitrary time interval $[t_0 - \Delta, t_0]$  where $\Delta$ is the size of the state estimation interval (typically 5 minutes in real time DTA systems). Assume that the length of the current state prediction horizon is equal to $H\Delta$ (each $\Delta$ interval within the prediction horizon is termed a prediction interval) and extends from  $[t_0, t_0 + H\Delta]$. In addition, assume that the link tolls are set for intervals of size $\Delta$ (this period is referred to as the tolling interval) and that the tolling intervals are aligned with the state estimation/prediction intervals. Let $ {\boldsymbol \tau^{h}} = (\tau^{h}_1 , \tau^{h}_2 \ldots \tau^{h}_{\widetilde{m}})$ represent the vector of link tolls for the time period $[t_0 + (h-1)\Delta , t_0 + h\Delta ]$ where $ h= 1 \ldots H$. The vector of tolls for the current prediction horizon is thus given by ${\boldsymbol \tau} = ({\boldsymbol \tau^1}
 , {\boldsymbol \tau^2 }, \ldots {\boldsymbol \tau^H} )$. 

In real world applications, given that the state estimation and solution of the optimization problem will require a finite computational time (assume that this is at most equal to the interval length $\Delta$), it will not be possible to implement the optimal toll vector for the first tolling interval within the prediction horizon.  
Consequently, the size of the optimization horizon is assumed to be one tolling interval less than the size of the prediction horizon and the decision variables in our optimization problem are in fact  ${\boldsymbol \tau^{'}} = ({\boldsymbol \tau^2 }, \ldots {\boldsymbol \tau^H} )$. ${\boldsymbol \tau^1 }$ is set to the optimal value for the same prediction interval from the previous execution cycle (denoted by ${\boldsymbol {\lambda}}$), so that ${\boldsymbol \tau} = ({\boldsymbol {\lambda}},{\boldsymbol \tau^{'}}) $. 

This is illustrated in the example in Figure \ref{fig:Formulation} for a case where $H=3$. In execution cycle 1 (denoted by C1), the decision vector consists of the toll values  $({\boldsymbol \tau^2_{C1} }, {\boldsymbol \tau^3_{C1}})$ for the prediction intervals P2 and P3. The toll vector ${\boldsymbol \tau^1_{C1} }$ is set as the optimal value from the previous execution cycle (denoted by ${\boldsymbol {\lambda}_1}$). Subsequently, in the second execution cycle, the decision vector consists of the toll values $({\boldsymbol \tau^2_{C2} }, {\boldsymbol \tau^3_{C2}})$ and ${\boldsymbol {\lambda}_2} = {\boldsymbol {\tau^{2*}_{C1} }}$, where $ {\boldsymbol {\tau^{2*}_{C1} }} $ is the optimal value of ${\boldsymbol \tau^2_{C1}}$ from execution cycle 1. 
 
Furthermore, consider the collection of vehicles $ \nu = 1, \ldots V$  on the network during the prediction horizon $[t_0, t_0 + H\Delta]$. Let the travel time of vehicle $ \nu $ be represented by $tt^{\nu}$ and the predictive travel time guidance be denoted by $ \mathbf{{tt^{g}}} = ( \mathbf{tt^{g}_i}; \forall i \in A )$, where $ \mathbf{tt^{g}_i}$ represents a vector of the time dependent link travel times (guidance) for link $i$. Note that the vehicle travel times $ \mathbf{tt} = (tt^{\nu} ; \nu = 1, \ldots V)$ are a result of the state prediction module of the DTA system and cannot be written as an explicit function of the tolls and predictive guidance. We characterize the complex relationship through a function $S(.)$ that represents the coupled demand and supply simulators as,
\begin{equation}
 S( \mathbf{x}^{p}, \boldsymbol{\upgamma}^{p}, \mathbf{tt}^{g}, {\boldsymbol \tau} ) = \mathbf{tt},    
\end{equation}

where $ \mathbf{x}^{p} ,\boldsymbol{\upgamma}^{p} $ represent the forecasted demand and supply parameters for the prediction horizon. Also note the iterative procedure described in Section  \ref{sec:Framework} ensures consistency between $ \mathbf{tt^{g}}$ and $ \mathbf{tt} $.

It is assumed that the total network demand is fixed (inelastic) and the behavioral response of users to the tolls and predictive travel time guidance is solely through route choice which is modeled within the demand simulator of DynaMIT2.0 using a path size logit model wherein the utility of a vehicle $\nu$ on path $k$  is given by, 
\begin{equation}
 \cup^{\nu}_{k} = \beta_{c}\:\widetilde{\tau}_{k} + \beta_{t}\:\bar{tt}^{g}_{k} + log(PS_{k}) + C_{k} + \epsilon^{\nu}_{k}, 
\end{equation}
where $ \widetilde{\tau}_{k} $ is the toll on route $k$ , $ \bar{tt}_{k}^{g}$ is the travel time on route $k$ as per the guidance information (which is the sum of travel times on component links), $\beta_c$ and $\beta_t$  represent the cost and travel time coefficients respectively, $PS_k$ represents the path size variable for path $k$, $C_k$  represents a composite utility pertaining to additional variables including path length, number of left turns and number of signalized intersections, $ \epsilon^{\nu}_{k}$ represents a random error term. Note that first, for vehicles that do not have access to the guidance information, historical travel times are used and second, similar model structures are used for both the pre-trip and en-route choice models. The reader is referred to \cite{BenAkiva_et_al:DynaMIT2010} for more details. 

It should be also be pointed out that since the optimization is performed within a rolling horizon framework and given that the tolls change every five minutes, it is likely that the toll values on which the driver based his pre-trip (or en-route) route choice decision are significantly different from the tolls he pays in reality. To mitigate the public opposition that may arise from this, we impose a limit on how much the tolls can vary across successive tolling intervals on a given gantry. Thus we have, 

\begin{equation}\label{eq:deltacon}
{\boldsymbol \tau^{h-1}} - {\boldsymbol \delta} \leq {\boldsymbol \tau^{h}} \leq {\boldsymbol \tau^{h-1}} + {\boldsymbol \delta }, \,\, h= 2, \ldots H,
\end{equation}
where ${\boldsymbol \delta } = (\delta_i; \forall i \in \widetilde{A} )$ represents the vector of limits on the change in tolls across successive intervals.  

With this background, the dynamic toll optimization problem in our context is formulated as a non-linear program in Equation \ref{eq:DTOP}. The objective function considered here is the total travel time of all vehicles on the network, but can be suitably modified to accommodate other objectives such as consumer surplus, operator revenues or social welfare depending on the context. The decision variables are the vector of toll values for the optimization horizon period. The constraints are the DTA system, upper and lower bounds on the toll values (denoted by vectors ${\boldsymbol \tau_{LB}}$ and ${\boldsymbol\tau_{UB}}$), and the constraints on changes in toll values across successive tolling intervals. 

\begin{equation}\label{eq:DTOP}
\begin{aligned}
  \bf{DTOP:} & \;\;\;\;\; \mathop{\textnormal{MIN}}_{{\boldsymbol \tau^{'}}}  \sum_{\nu=1}^{V} tt^{\nu}({\boldsymbol \tau^{'}}) \\  
  & s.t. \\
  & S( \mathbf{x}^{p}, \boldsymbol{\upgamma}^{p}, \mathbf{tt}^{g}, {\boldsymbol \tau}) = \mathbf{tt}, \\ 
  & {\boldsymbol \tau^{h-1}} - {\boldsymbol \delta} \leq {\boldsymbol \tau^{h}} \leq {\boldsymbol \tau^{h-1}} + {\boldsymbol \delta }, \,\, h= 2, \ldots H, \\
  & {\boldsymbol \tau_{LB}} \leq {\boldsymbol\tau^{h}} \leq {\boldsymbol\tau_{UB}},\; h=2, \ldots H. \\ 
\end{aligned}
\end{equation}

In case of computational performance constraints, the dimensionality of the \textbf{DTOP} problem above may be significantly reduced by assuming that the vector of tolls does not change across prediction intervals within the optimization horizon. In other words, we assume that $ ({\boldsymbol \tau^2 } = {\boldsymbol \tau^3 } \ldots =  {\boldsymbol \tau^H} =\bar{ {\boldsymbol \tau}} )$ which reduces the number of decision variables from $ \widetilde{m}(H-1) $ to $ \widetilde{m} $. In this case, the constraints defined by Equation \ref{eq:deltacon} are replaced by, 

\begin{equation}\label{eq:deltacon_2}
{\boldsymbol \lambda} - {\boldsymbol \delta} \leq \bar{ {\boldsymbol \tau}} \leq {\boldsymbol \lambda} + {\boldsymbol \delta }
\end{equation}

\section{Solution Algorithm} \label{sec:algorithm}
As noted earlier, since the objective function of the dynamic toll optimization problem in our context does not have a closed form and is the output of a complex simulator, evolutionary algorithms and meta-heuristics are preferable to classical gradient based approaches. Hence, a real-coded Genetic Algorithm (GA) \cite{nsgaii} is applied to solve the DTOP problem formulated in Section \ref{sec:Formulation}.

The algorithm starts by randomly generating a set of control strategies or individuals ($CS^p_i, i=1 \ldots N, p:parent$) collectively known as the initial population or parent population of size $N$. Each single control strategy $CS^p_i$ comprises the vector of tolls ${\boldsymbol \tau^{'}} = ({\boldsymbol \tau^2 }, \ldots {\boldsymbol \tau^H} )$ for the current optimization horizon. These $N$ different control strategies $CS^p_{i},\; i=1 \ldots N $ (or individuals) are evaluated (i.e the objective function value is computed) in parallel by running the state prediction module of DynaMIT2.0 ($ PD_i, \; i=1 \ldots N $) independently for each control strategy $CS^p_i$. The modularity of DynaMIT2.0 provides the functionality to execute a single state estimation run followed by multiple parallel state predictions with different control strategies and makes real time optimization (within the budget of 5 minutes) computationally possible.

Different control strategies (or individuals) in the parent population are assigned a rank $ R_i,\; i=1 \ldots N$ based on their respective objective function values $ Obj_i,\; i=1 \ldots N $. The objective function value for each individual is calculated from the output of the state prediction module $PD_i$. From this set of control strategies (individuals) in the parent population, a new set of control strategies (new individuals), collectively called the child population of size $N$ is generated using genetic operators, i.e., the SBX crossover and polynomial mutation.  The newly generated $N$ control strategies (or individuals) of the child population ($CS^c_i, i=1 \ldots N, c:child$) are evaluated in parallel by running the state prediction module (on the same estimated state used to evaluate parent population) to get their objective function values.

Strategies in the child population $CS^c_i, i=1 \ldots N$ and parent population $CS^p_i, i=1 \ldots N$ are merged together to form a combined set of strategies (mixed population) of size $2N$, $CS^{p+c}_k (k=1 \ldots 2N)$  and are ranked based on their objective function values. From these $2N$ different strategies (or individuals) $CS^{p+c}_k$, the best $N$ strategies (or individuals) are selected based on their rank to form the parent population for the next iteration. The procedure of generating a new set of strategies (or child population) continues uptill the termination criteria are met. The termination criteria can be a predefined number of iterations/generations $G_{max}$, a threshold for the improvement in the objective function value or a computational time budget $T_{max}$. 

\begin{figure}[h]
\centering 
{\includegraphics[width=0.49\textwidth]{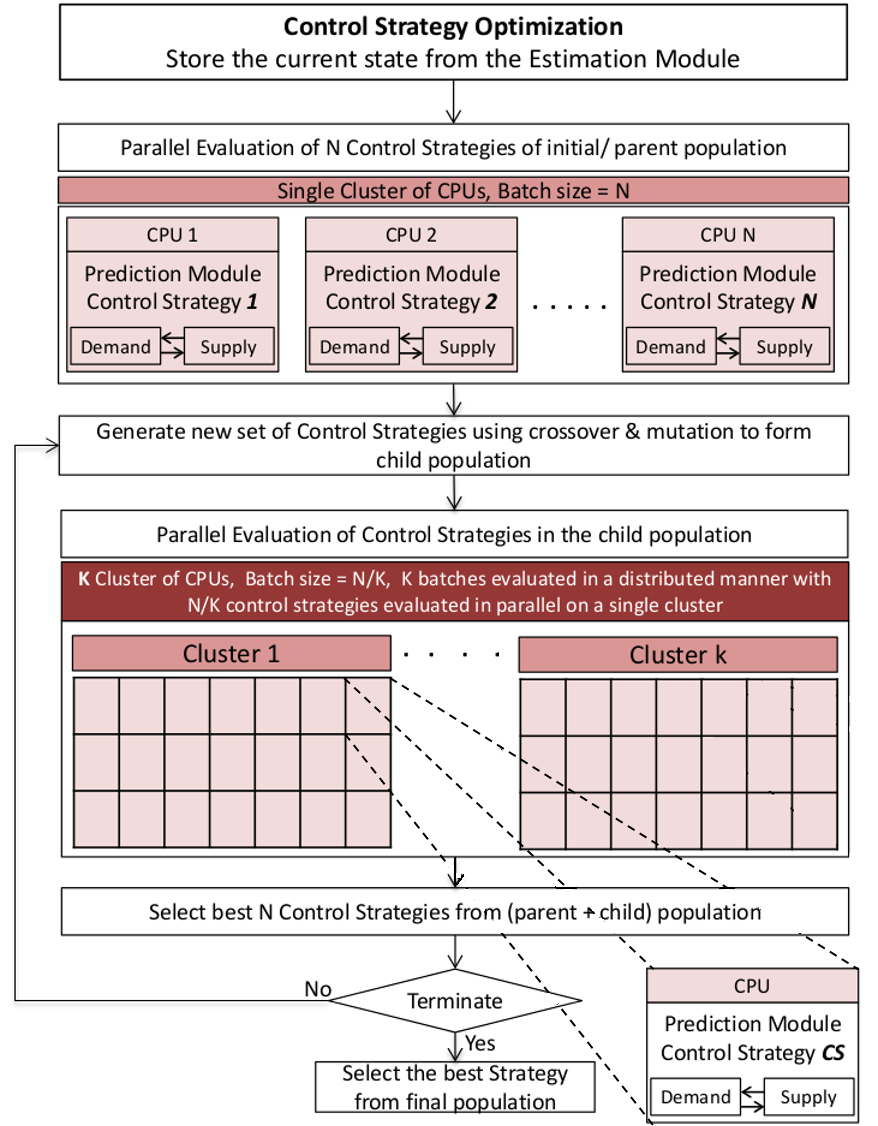}}
\caption{Genetic Algorithm with parallel evaluation of population using parallel \& distributed computing techniques.}
\label{fig:flowchart}
\end{figure}

In order to facilitate \textit{real time} performance, it is imperative to compute parallel tasks in a computationally efficient manner using a multi-core architecture. In the context of our framework, evaluation of different control strategies $CS_i$s in a particular iteration are independent of each other, therefore evaluating them in parallel significantly reduces computational time and makes the approach scalable. Different parallel computing architectures and libraries have been proposed in the literature, but for our application we adopt a Master-Slave architecture using the GNU\footnote{GNU is a recursive acronym for \textbf{G}NU's \textbf{N}ot \textbf{U}nix.} Parallel library \cite{parallel}. The main benefit of using GNU Parallel is that processor level parallelism can be achieved. To evaluate each control strategy, a new process is launched on a different CPU. Managing and scheduling of different processes is taken care by the GNU library at the operating system level. In practice, it is observed that even with processor level parallelism the speed-ups are not linear even if no inter-process communication is present. Therefore, the framework is designed so as to allow \textit{Batch-Wise} evaluation of different control strategies. Specifically, during each iteration, all $N$ different control strategies can be launched as different processes on different CPUs, or they can be launched in batches of size $n, n < N$. In this batch wise implementation, different batches can either be launched  sequentially on a single cluster of CPUs or they can even be launched in parallel on multiple clusters of CPUs. This framework exploits both parallel as well as distributed computing simultaneously as shown in Figure \ref{fig:flowchart}.

\begin{algorithm}
\caption{Control Strategy Optimization Algorithm}\label{fill_nds}
   \begin{enumerate}
      \item Store the output of the \textit{State Estimation} module for interval $[t_0 - \Delta, t_0]$
      \item Replicate the cached estimated state to $N$ different DynaMIT clones 
      \item Initialize $N$ different control strategies to form initial population
      \item Evaluate $N$ different control strategies by running \textit{State Prediction} module iteratively for interval $[t_0, t_0 + H\Delta]$ using $N$ different DynaMIT clones in parallel
      \item Assign rank to $N$ individuals on the basis of their Objective Function value.
 \begin{algorithmic}[1]
      \ForAll{$g \gets 2$ to $G_{max}$ $\cap$ clock-time $\leq T_{max} $}
	\begin{enumerate}
	 \item Generate $N$ new control strategies using tournament selection with SBX crossover to form child population
	 \item Mutate $N$ newly generated control strategies using polynomial muatation
	 \item Evaluate $N$ child strategies by running \textit{State Prediction} module for interval $[t_0, t_0 + H\Delta]$ using $N$ different DynaMIT clones in parallel
	 \item Merge the child population and parent population strategies to form mixed population of size $2N$
	 \item Assign rank to $2N$ individuals on the basis of their Objctive Function value
	 \item Select the best $N$ individuals (strategies) to form the parent population
	\end{enumerate}
      \EndFor
 \end{algorithmic}
  \item Select the best Control Strategy $CS_{best}$ from the final set of strategies
  \end{enumerate}
\end{algorithm}

\section{Experiments}\label{sec:Experiments}
\begin{figure}[t]
\centering
\includegraphics[width=0.47\textwidth]{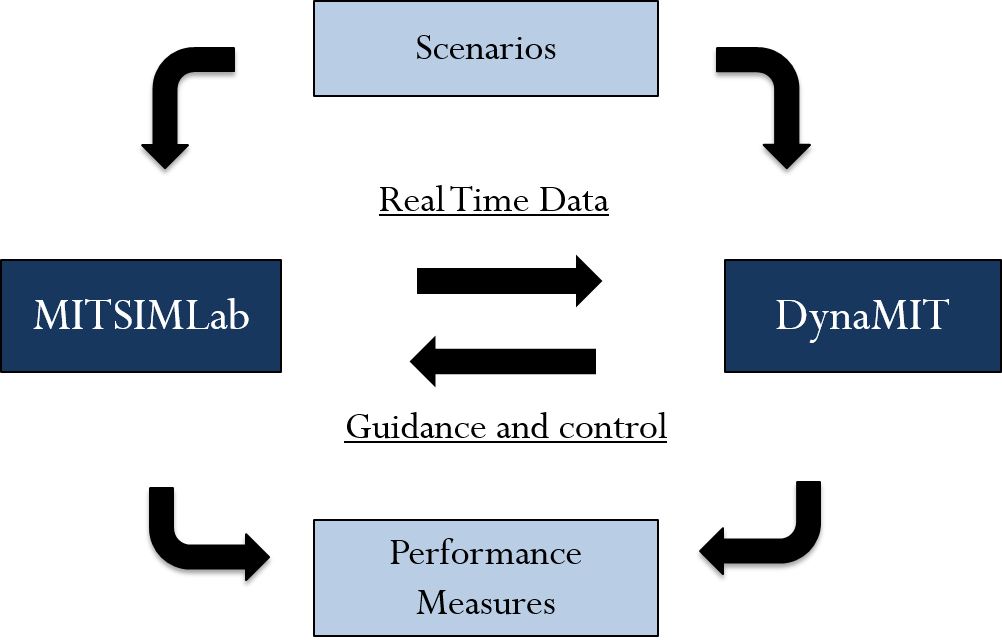}
\caption{Closed-Loop Framework}
\label{fig:ClosedLoop}
\end{figure}

This section discusses results from a set of experiments conducted to investigate the performance of the proposed strategy optimization approach using DynaMIT2.0 on the Singapore expressway network. The numerical experiments are conducted using a closed-loop framework, interfacing DynaMIT2.0 and MITSIMLab (MITSIM), a microscopic simulator \cite{BenAkiva_et_al:MITSIMLab2010}. MITSIM is run concurrently with DynaMIT and mimics the real network, providing sensor counts for the current interval to DynaMIT which in turn provides predictive guidance and tolls to MITSIM (see Figure \ref{fig:ClosedLoop}). The effect of the guidance and tolls can then be examined by extracting relevant performance measures from MITSIM avoiding overestimation of the benefits. 

\begin{figure}[h]
\centering
\includegraphics[width=0.475\textwidth]{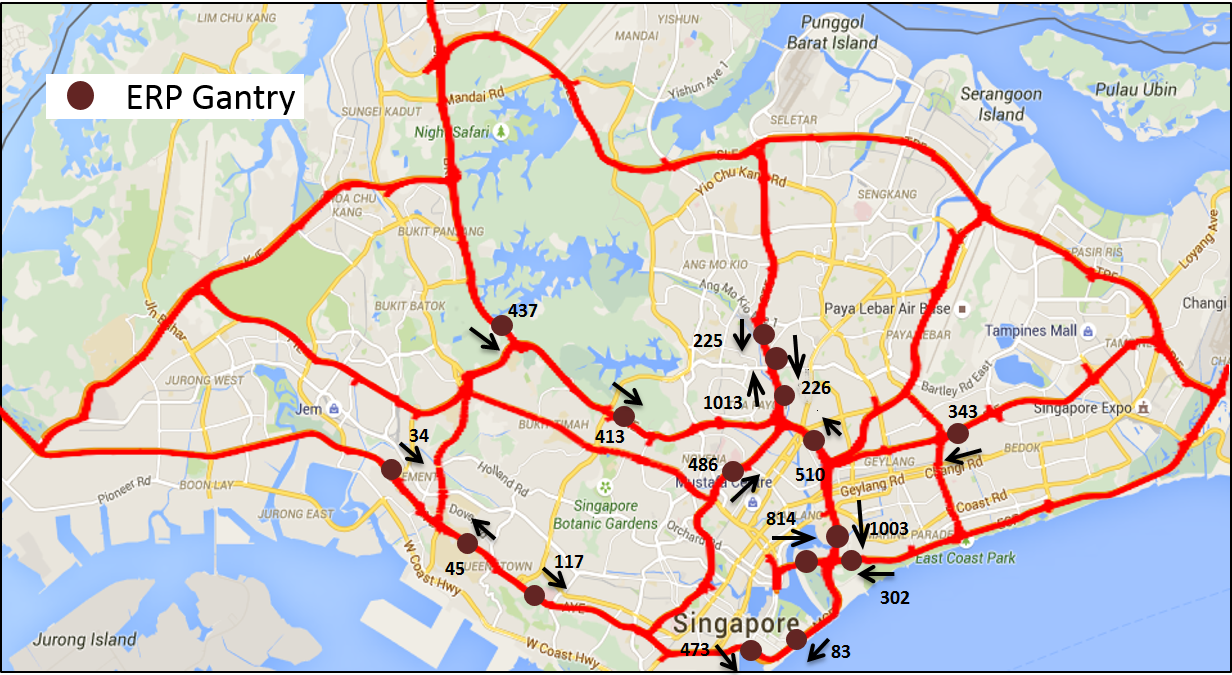}
\caption{Network of Expressways and Major Arterials in Singapore}
\label{fig:Network}
\end{figure}

The experiments are conducted on the network of major arterials and expressways in Singapore (Figure \ref{fig:Network}) which consists of 948 nodes, 1150 links, 3891 segments, and 4123 origin-destination (OD) pairs, and 16 tolled links. The labels represent the links where there is a toll gantry.

The section is organized into five parts. The first part discusses the setup of the closed-loop framework and calibration, the second describes the experimental design and inputs. The third section analyzes the results in terms of travel time savings and the effect of network demand, fourth part discusses the optimal tolls through few gantries, and finally the fifth part discusses computational performance.  


\subsection{Closed-Loop Calibration}\label{sec:CLcalib}
In order to set up the closed-loop environment, a two stage calibration procedure is adopted using the w-SPSA algorithm \cite{lu2014wspsa} (for other approaches see \cite{ciuffo2014}). In the first stage, dynamic OD demand (for a period between 06:30 AM and 12:00 PM), driver behavior and route choice parameters of MITSIM are calibrated by minimizing a two component objective function. The first component is the sum of squared deviations between simulated counts and actual counts (on a set of 325 sensors for 5 minute time intervals averaged across 30 weekdays in February and March 2015) obtained from the Singapore Land Transport Authority (LTA). The second component is the difference between the parameter values and apriori estimates. The inputs for the calibration process is a set of a priori parameter values and a seed OD matrix obtained from a prior calibration procedure \cite{lu2014wspsa}. The normalized root mean square error in the sensor counts before and after the calibration process were 73\% and 34\% respectively.

\begin{figure*}[ht]
\centering
\subfloat[Simulated vs Actual Count \label{fig:ClosedLoopCalibration} ]{\includegraphics[height=50mm, width=0.40\textwidth]{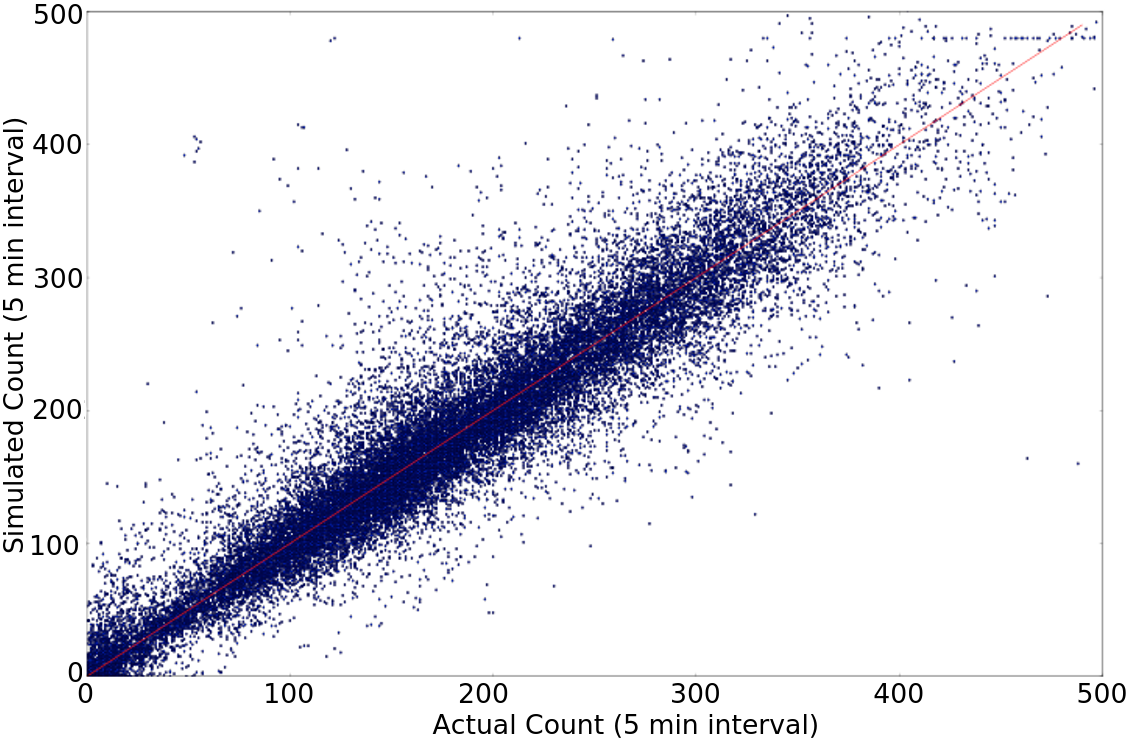}}
\subfloat[Simulated vs Actual Travel Times \label{fig:ClosedLoopCalibration_TT}]{\includegraphics[height=50mm, width=0.40\textwidth]{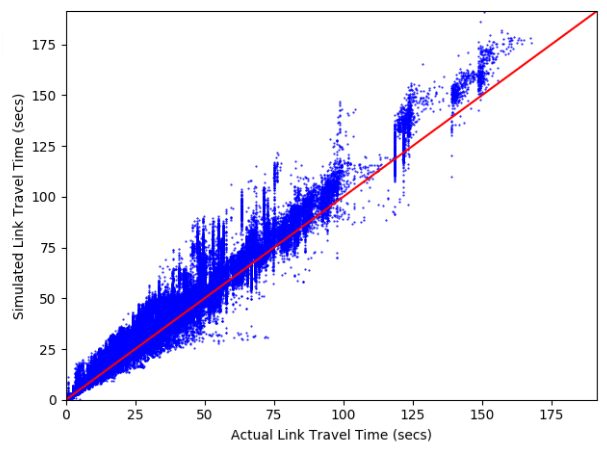}}
\caption{Closed Loop Calibration}
\end{figure*}

In the second stage, the historical OD matrix, supply and route choice parameters of DynaMIT2.0 are calibrated against the outputs (sensor counts on 650 network segments) generated by MITSIM. The normalized root mean square error in the sensor counts before and after the calibration process were 56\% and 19\% respectively. Further, the RMSN in time-dependent link travel times after calibration was found to be 24\%. The results of the second stage of the calibration are summarized in Figures \ref{fig:ClosedLoopCalibration} and \ref{fig:ClosedLoopCalibration_TT} which show scatter plots of the simulated (DynaMIT) versus actual (MITSIM) sensor counts and link travel times respectively.

\subsection{Experimental Setup}\label{sec:setup}
The numerical experiments are conducted using a simulation period from 6:30 AM to 12:00 PM which includes the morning peak in Singapore. The state estimation interval (and OD demand interval) is five minutes ($\Delta = 300$ seconds) and the prediction horizon is 15 minutes ($H=3$). The simulation period is composed of three parts: a \textit{Warm-up} period from 6:30-7:30 AM where no tolls are imposed, a \textit{tolling period} from 7:30 - 11:00 AM, and a \textit{post-tolling} period from 11:00 AM to 12:00 PM where again no tolls are imposed.  

The impact of the predictive toll optimization is examined against two benchmarks using the closed-loop framework described earlier. It is assumed that the base demand (MITSIM OD demand obtained from the closed-loop calibration) represents the historical demand or an "average" day. This demand is then perturbed to reflect day to day variability by sampling from a normal distribution with expected value as the base demand and a coefficient of variation of 0.2. 
%
%

The first benchmark is the \textit{\textbf{no toll}} scenario where the closed-loop is simulated using the perturbed demand with zero tolls. The second scenario consists of \textit{\textbf{static optimized tolls}}. In this scenario, we first compute the optimum static tolls which involves minimizing the total travel times for the entire simulation period (obtained from the state estimation) by implementing a single vector of tolls for the complete tolling period. The closed-loop is now simulated using the perturbed demand with the static optimum tolls. Finally, in the third scenario the closed-loop is simulated using the perturbed demand and the \textit{\textbf{predictive optimized tolls}} based on the proposed framework in Section \ref{sec:Framework}. In all three scenarios, MITSIM receives predictive travel time guidance from DynaMIT2.0 and in turn provides sensor counts to DynaMIT2.0 every estimation interval (or execution cycle). 

Further, to investigate the effect of the overall demand level, all the three aforementioned scenarios are simulated for four different demand levels: \textit{low} (base demand reduced by 10\%), base (closed-loop calibration as noted earlier) , high (base demand increased by 10\%) and \textit{very high} (base demand increased by 20\%). Note that the demands referred to here are the actual MITSIM (real world) demands. 
For the scenarios with predictive optimization, the DynaMIT2.0 historical
demand (obtained from the second stage in the closedloop
calibration) remains unchanged for all demand scenarios. For the scenarios with the static optimum tolls, note that the regulator must perform the determination of the optimum tolls 'offline' using an estimate of historical demand. Given that different levels of actual demand (unknown to the regulator) are tested, we assume that a single computation of the static optimum tolls is performed by considering a worst case scenario where the calibrated DynaMIT2.0 historical demand is increased by 20\%. In addition to the comparison with predictive optimization, this allows us to also test the robustness of the static optimum tolls to both systematic and random variation in the actual OD demands (from historical estimates). 

The performance measures are: 1) average travel times (across vehicles) for each departure time interval obtained from MITSIM, 2)  computational time for each execution cycle of DynaMIT2.0. 
Note that for each scenario and demand level, the performance measures reported are averages across 10 different runs to account for stochasticity in the overall system.

A High Performance Computing Cluster (HPCC) with 120 CPUs and 256 GB of memory is used to run the experiments. For the parameters of GA, we use a population size of 60, probability of cross-over and mutation as 0.7 and 0.1 respectively with a computation budget of 300 seconds. The number of iterations may vary from interval to interval depending on the demand, i.e., peak or off-peak periods.

\subsection{Analysis of Travel Time} \label{sec:analysis}
\begin{SCfigure*}
\caption{Travel Time for Low Demand}
\includegraphics[height= 61mm, width=0.8\textwidth]{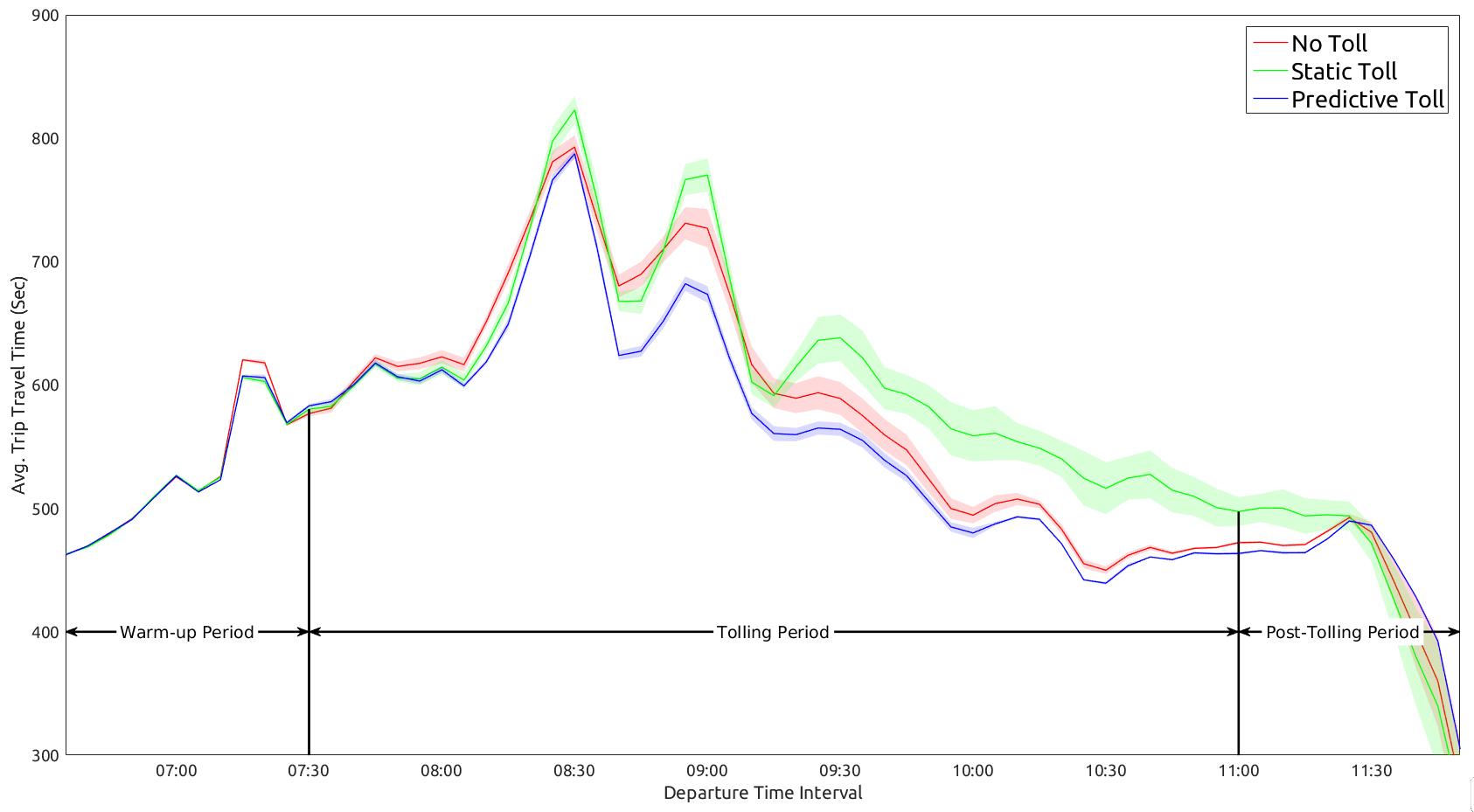}\label{fig:TT_dem6}\\[-5ex]
\end{SCfigure*}

\begin{SCfigure*}
\caption{Travel Time for Base Demand}
\includegraphics[height= 61mm, width=0.8\textwidth]{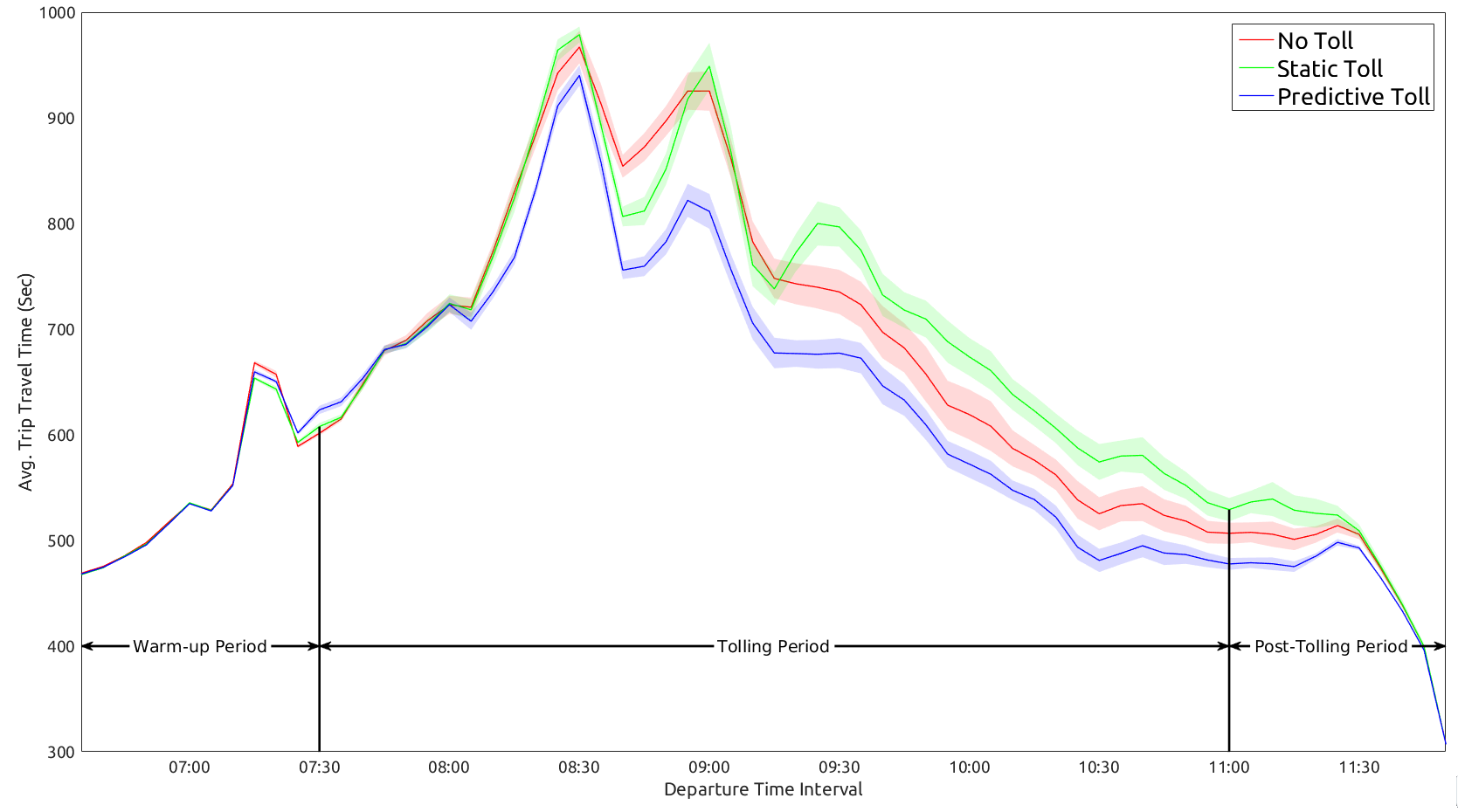}\label{fig:TT_base}\\[-5ex]
\end{SCfigure*}

\begin{SCfigure*}
\caption{Travel Time for High Demand}
\includegraphics[height= 61mm, width=0.8\textwidth]{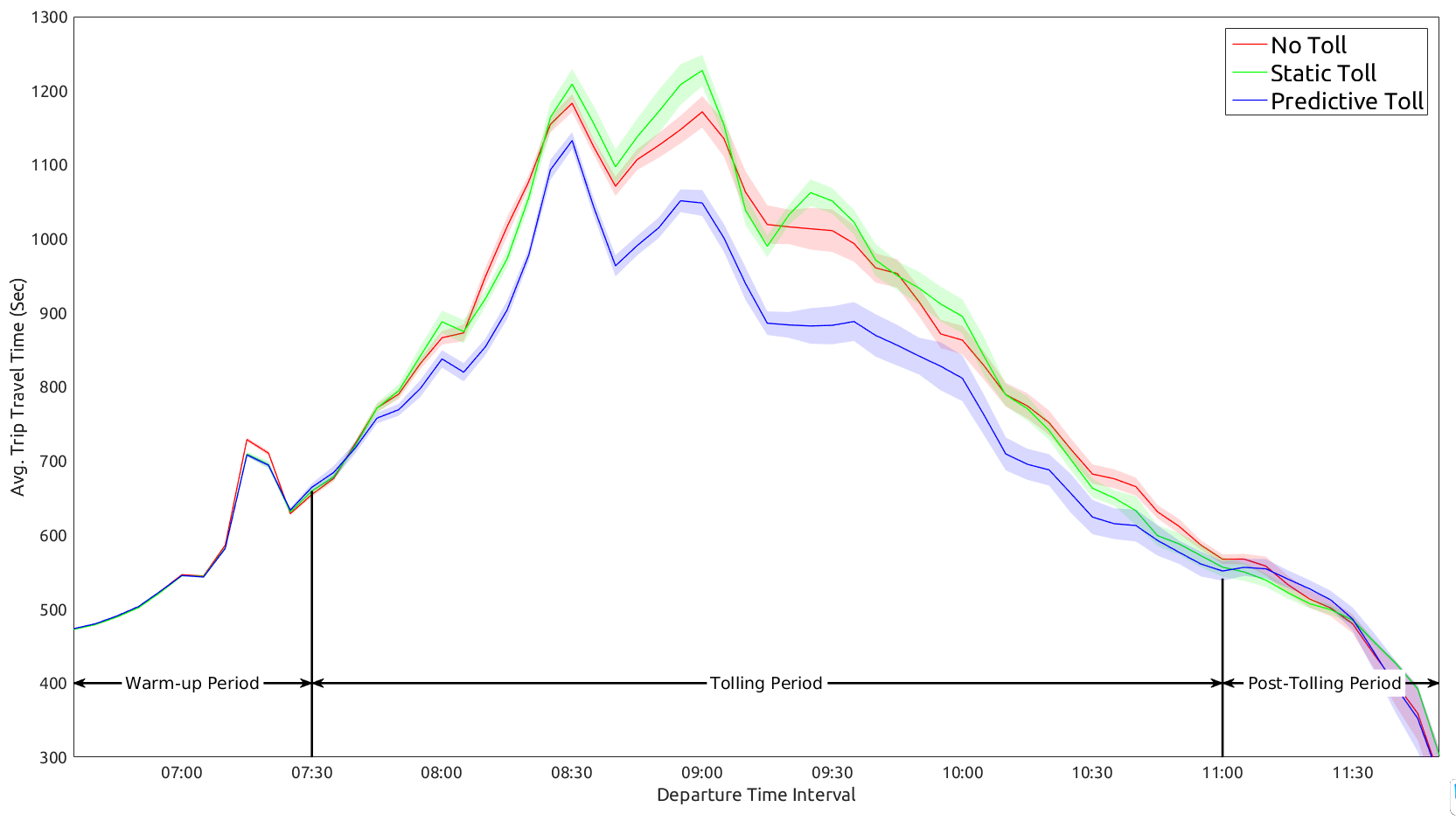}\label{fig:TT_dem7}\\[-5ex]
\end{SCfigure*}

\begin{SCfigure*}
\caption{Travel Time for Very High Demand}
\includegraphics[height= 61mm, width=0.8\textwidth]{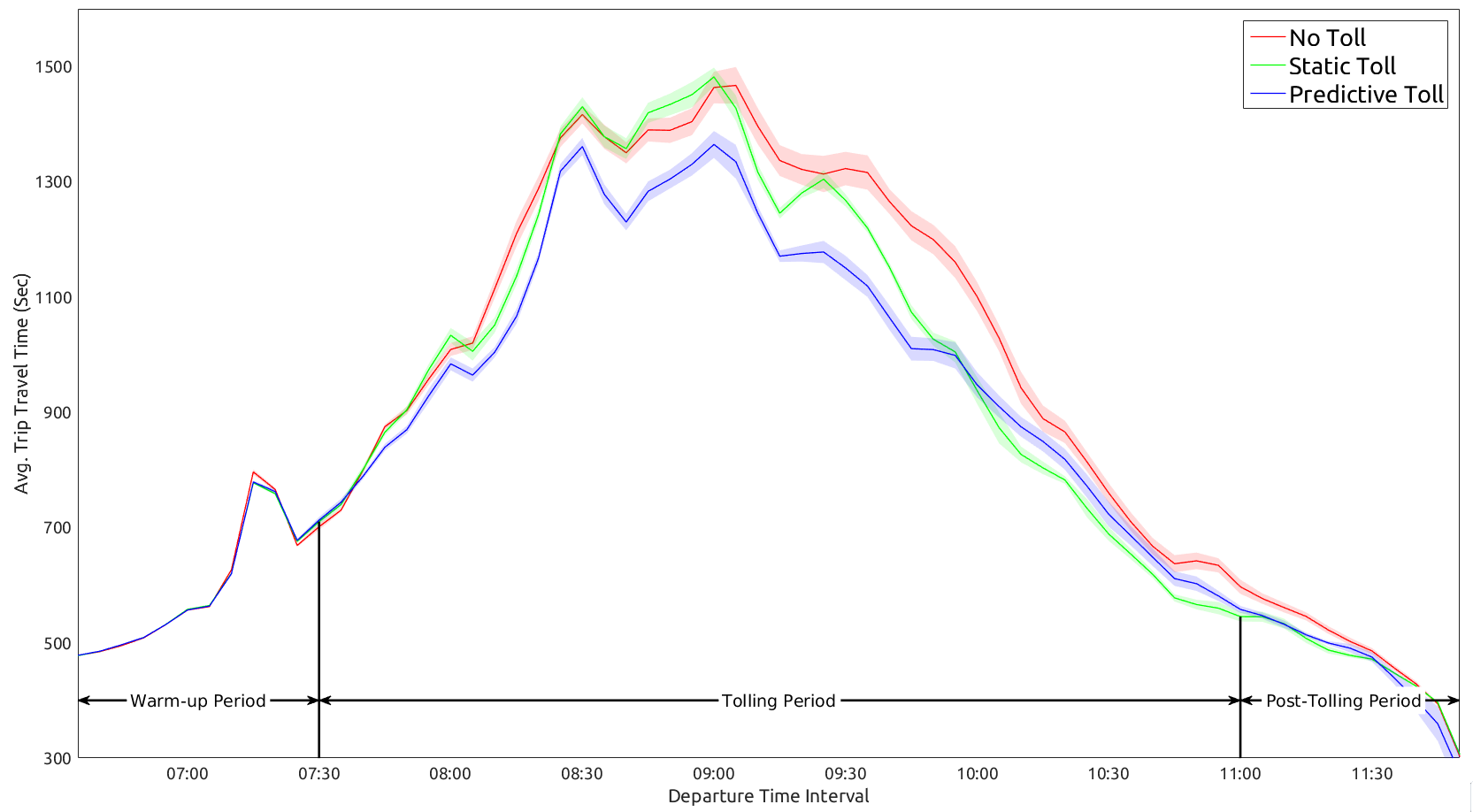}\label{fig:TT_vhigh}
\end{SCfigure*}

In order to compute and compare average time-dependent travel times across scenarios, for the entire population, all the drivers departing in a given time interval (e.g., 07:00-07:05) are identified and their average trip travel time is calculated. This process is repeated for each consecutive 5 min interval in the entire simulation period, i.e., starting from 6:30-6:35, 6:35-6:40, ...., up to 12:25 -12:30. 
The results indicate that the use of predictive optimized tolls yields significant travel time savings over both the no toll and static optimum scenarios. The percentage improvement in travel times of the predictive optimized toll scenarios over the two benchmark scenarios for the tolling period and peak period (for all demand levels) is summarized in Table \ref{tab:summary}. The average travel times (over the tolling period) in the case of the predictive optimized tolls are lower than the static optimum and no toll cases by 9.12\% and 6.74\% in the base demand case. Interestingly, the static optimum is worse than the no toll case for the low, base and high demand scenarios (see also Figures \ref{fig:TT_dem6} to \ref{fig:TT_vhigh}). This indicates that the static optimum based on historical demands is not robust when the actual demands vary significantly from the historical estimates. Note that the historical demand was scaled up by 20\% when computing the static optimum and hence, in the very high demand case where the historical estimates are closest to the actual demands, the static outperforms the no toll scenario. The  percentage decrease in travel time is 5.39\% and 3.71\% in the low demand case, 8.88\% and 8.24\% in the high demand case and 4.00\% and 8.38\% in the very high demand case. In addition, the percentage improvement for the peak period (between 8:00 am and 9:30 am) is 7.94\% with respect to the no toll scenario and 8.36\% with respect to the static optimum scenario for the base demand case. It should be noted that in event of non-recurrent scenarios (like a special event or an incident) one would expect a significantly higher impact of the toll optimization and guidance provision. 

Furthermore, for all demand cases, a standard two sided t-test indicates that the mean travel time (for all departure time intervals within the peak period) of the predictive optimized tolling scenario has a statistically significant difference from that of the no toll/static optimum scenarios at a confidence level of $\alpha=95$\%. 

\sidecaptionvpos{figure}{h}
 
\begin{figure*}[ht]
\centering{
\subfloat[Base Demand pdf]{\includegraphics[height=50mm, width=0.40\textwidth]{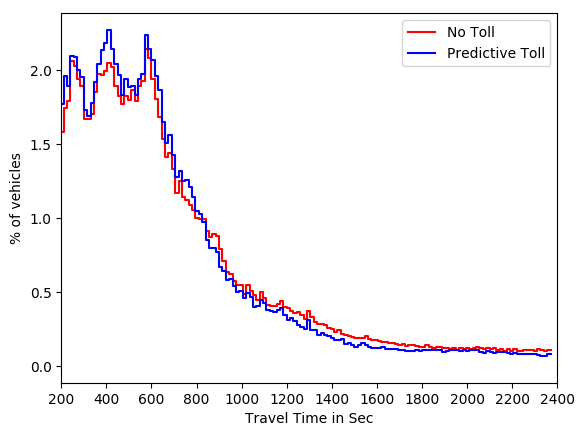}}
\subfloat[Base Demand cdf]{\includegraphics[height=50mm, width=0.40\textwidth]{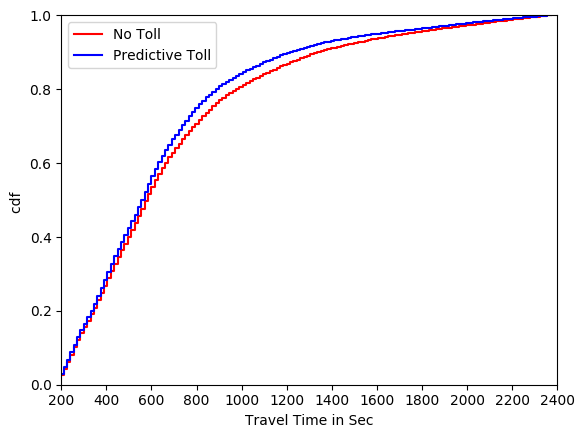}}
\\[-2ex]
 \subfloat[High Demand pdf]{\includegraphics[height=50mm, width=0.40\textwidth]{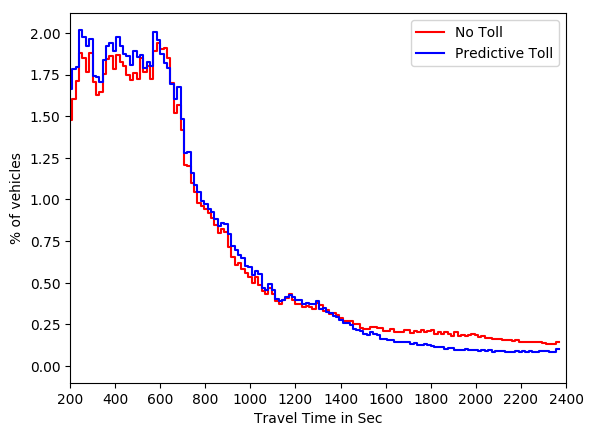}}
 \subfloat[High Demand cdf]{\includegraphics[height=50mm, width=0.40\textwidth]{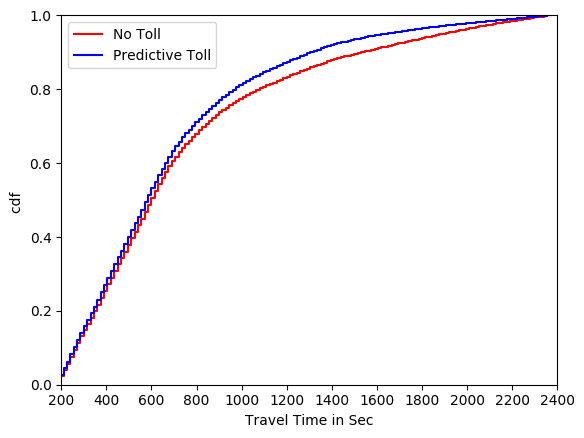}}
\caption{\label{fig:TTDist} Peak Period Travel Time Distributions}
}
\end{figure*}

Figures \ref{fig:TT_dem6} to \ref{fig:TT_vhigh} plot the mean travel times (shaded region represents the standard error in estimate of the mean) versus departure time interval for the three scenarios and each demand level. With regard to the effect of the overall demand level on the improvement in travel time savings with respect to the static/no-toll toll case, the results indicate the lowest improvements (during the peak period) are attained when the congestion levels are either very low or very high. This occurs because in the ’low’ demand scenario the relatively uncongested state of the network reduces the impact of toll optimization. On the other hand, the severely congested network state in the very high demand scenario also reduces the possibility of alleviating congestion through the re-routing of vehicles leading once again to smaller benefits of the toll optimization.



The probability density and cumulative density functions of vehicle travel time are plotted in Figure \ref{fig:TTDist} for the no toll and predictive optimized toll scenarios in the base and high demand cases. The plots highlight the reduction in frequency of trips with higher travel time due to the predictive optimization of tolls.

\begin{table}[t]
\centering
\caption{Travel Time Improvement}
\begin{tabular}{@{}|c|c|c|c|c|c|@{}}
\toprule
\multirow{3}{*}{Demand Level} & \multicolumn{4}{c|}{\% Travel Time improvement}                        & \multirow{3}{*}{\begin{tabular}[c]{@{}c@{}}Total Drivers \\ Simulated\end{tabular}} \\ \cmidrule(lr){2-5}
                              & \multicolumn{2}{c|}{Tolling Period} & \multicolumn{2}{c|}{Peak Period} &                                                                                     \\ \cmidrule(lr){2-5}
                              & No Toll           & Static          & No Toll         & Static         &                                                                                     \\ \midrule
Low                           & 3.71               & 5.39            & 7.61            & 6.25           & 275,000                                                                             \\ \midrule
Base                          & 6.74             & 9.12            & 8.36           & 7.94           & 300,000                                                                             \\ \midrule
High                          & 8.24             & 8.88             & 9.65           & 10.74           & 325,000                                                                             \\ \midrule
Very High                     & 8.38              & 4.00            & 8.20            & 7.01           & 350,000                                                                             \\ \bottomrule
\end{tabular}
\label{tab:summary}
\end{table}

\subsection{Analysis of Optimized Tolls} \label{sec:optTolls}
Here we provide few examples in order to analyze the optimized tolls under predictive optimization with respect to static optimization. 

\begin{figure}[H]
\centering{
 \includegraphics[height =30mm]{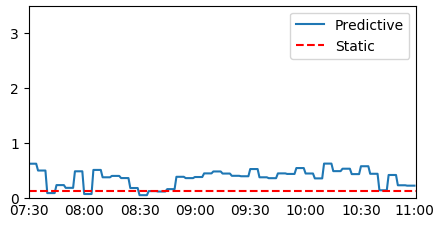}
 \caption{Gantry on link 45}\label{fig:gantry45}
 \includegraphics[height =30mm]{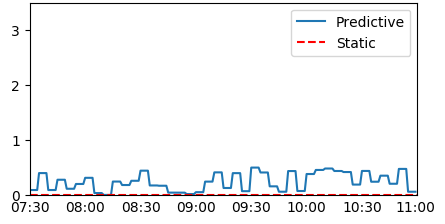}
 \caption{Gantry on link 83}\label{fig:gantry83}
  \includegraphics[height =30mm]{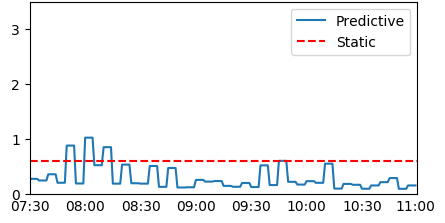}
 \caption{Gantry on link 225}\label{fig:gantry225}
 \includegraphics[height =30mm]{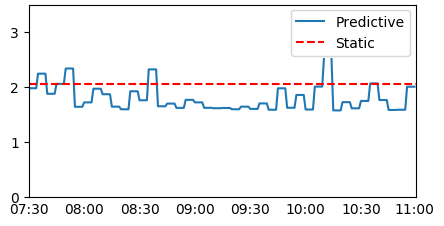}
 \caption{Gantry on link 226}\label{fig:gantry226}
 }
\end{figure}

First, we give an example of two gantries on links 45 and 83. We present the optimized tolls under static and predictive strategies in Figure \ref{fig:gantry45} and \ref{fig:gantry83}. The most preferred path for one of the ODs with a very high demand during the morning peak uses these gantries (first 83 and then 45). The predictive tolls are optimized at higher values compared to the static case and this indicates that real-time predictive tolls are adjusted better with respect to demand. 
Second, gantries on links 225 and 226 are optimized at lower values during the peak compared to static strategy as shown in Figure \ref{fig:gantry225} and \ref{fig:gantry226}.  It is observed that these gantries are used towards destinations that have very low demand in the morning peak. Predictive toll optimization is able to lower the tolls during the peak in order to account for lower demand values towards better travel times. 

\subsection{Computational Performance}
\begin{figure*}
\centering
\includegraphics[scale=0.35]{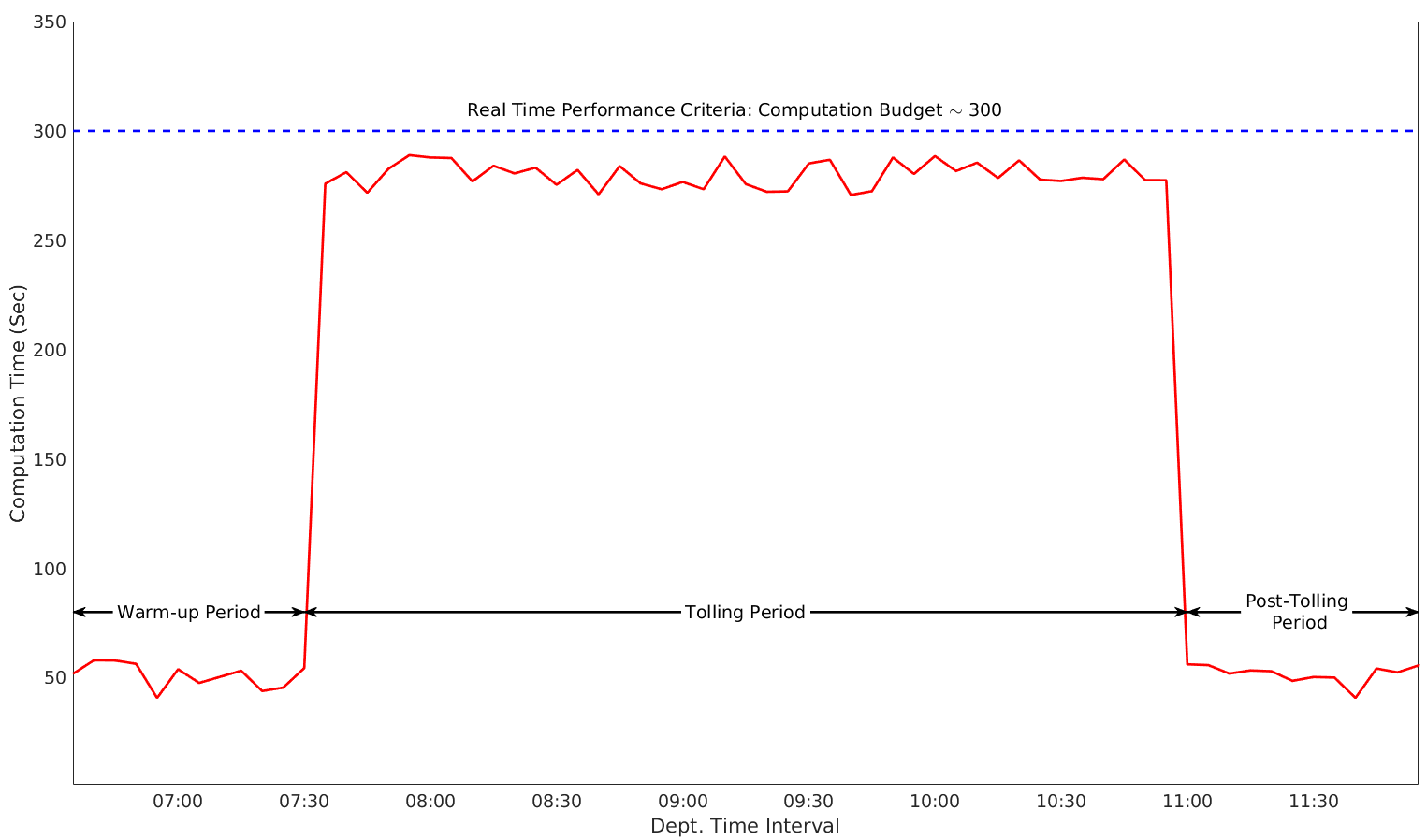}
\caption{Computational Performance}\label{fig:CompTime}
\end{figure*}
The results also indicate that the proposed solution algorithm achieves real-time performance, i.e the average computational time per execution cycle (across all demand levels) is within the five minute time budget (less than a single state estimation interval) discussed in Section \ref{sec:Formulation}. The plot of average computational time versus time interval is shown in Figure \ref{fig:CompTime}. 

The tractable computational times are the result of three contributing factors. The first is the imposition of the constraint on the extent to which tolls on a given gantry can vary across successive tolling intervals which significantly reduces the search space for the GA. This ensures that a population size of 60 suffices to attain a significant reduction in travel times within a low computational time budget. Secondly, the rolling horizon approach implies that the system is re-optimized every five minutes and consequently a poor solution in one interval can be quickly rectified or improved in subsequent intervals. This along with the feedback from the real network to the DTA system (through the online calibration) makes the control strategy optimization framework more robust. Finally and most importantly, the synchronous parallel evaluation of strategies in each iteration of the optimization procedure allows for evaluation of a sufficiently large number of candidate solutions. 


\section{Conclusion}
This paper proposes an integrated framework that combines the optimization of network control strategies with the generation of consistent guidance information for real-time DTA systems. The efficacy of the  proposed framework is demonstrated through a fixed demand dynamic toll optimization problem. Furthermore, a highly parallelizable genetic algorithm based solution approach is adopted. Numerical experiments conducted on a large scale real world network (expressways and major arterials in Singapore) indicate that use of the proposed framework can yield significant network-wide travel time savings of up to 8.36\% and 7.94\% over the no toll and static optimum scenarios respectively. A sensitivity analysis of demand levels further indicate that the highest improvements are attained at moderate and high demand levels. Finally, the proposed solution algorithm achieves real-time performance with a computational time of less than 5 minutes for each execution cycle within the rolling horizon scheme. The proposed framework and solution approach have important applications for real-time traffic management and advanced traveler information systems.

Some directions for future research include the application of the strategy optimization framework under non-recurrent scenarios including crisis-management \cite{HETU201815}, consideration of other objectives such as consumer surplus, operator revenue and multiple objectives; incorporation of traffic state prediction errors \cite{pereira2014} and the modeling of elastic demand through trip cancellation and departure time shifts in response to tolls. The application to other network control strategies and examination of the suitability of alternative solution algorithms also promise to be interesting areas for future research.



\section*{Acknowledgment}
We would like to thank the Singapore Land Transport Authority (LTA) ITS division, for providing the data for this project. This work was supported by the National Research Foundation of Singapore (SMART program).

\ifCLASSOPTIONcaptionsoff
  \newpage
\fi



%
\bibliographystyle{IEEEtran}
\bibliography{references}

\end{document}